\documentclass{elsart1p}

\usepackage[pdftex]{graphicx}

\usepackage{amssymb}

\def\nc{\newcommand}
\nc{\shalf}{\ensuremath{\textstyle \frac{1}{2}}}
\nc{\deldag}{\mathbin{\partial\mkern-10.5mu\big/}}
\nc{\deldagss}{\mathbin{\partial\mkern-10.5mu/}}
\nc{\kdag}{\mathbin{k\mkern-10mu\big/}}
\nc{\udag}{\mathbin{u\mkern-10mu\big/}}
\nc{\kdagss}{\mathbin{k\mkern-10mu/}}
\nc{\Pdag}{\mathbin{P\mkern-10mu\big/}}
\nc{\pp}{{\scriptscriptstyle ||}}
\nc{\stwo}{{\scriptscriptstyle 2}}
\nc{\pham}{{\phantom{-}}}

\def\lsim{\mathrel{\raise.3ex\hbox{$<$\kern-.75em\lower1ex\hbox{$\sim$}}}}
\def\gsim{\mathrel{\raise.3ex\hbox{$>$\kern-.75em\lower1ex\hbox{$\sim$}}}}

\def\Slashnew#1{#1\kern-0.55em\raise.05ex\hbox{/}}
\def\slashnew#1{#1\kern-0.5em\raise.05ex\hbox{{$\scriptstyle /$}}}
\def\emph#1{{\em #1}}

\nc{\beq} {\begin{equation}}
\nc{\eeq} {\end{equation}}
\nc{\beqa}{\begin{eqnarray}}
\nc{\eeqa}{\end{eqnarray}}

\begin{document}

\begin{frontmatter}



\title{Kinetic transport theory with quantum coherence}


\author{Matti Herranen, Kimmo Kainulainen and Pyry M.~Rahkila}

\address{University of Jyv\"askyl\"a, Department of Physics,\\ 
        P.O.~Box 35 (YFL), FIN-40014 University of Jyv\"askyl\"a, Finland \\
        and \\
        Helsinki Institute of Physics, P.O.~Box 64, FIN-00014 University of  		
   	    Helsinki, Finland.}

\begin{abstract}
We derive transport equations for fermions and bosons in spatially or temporally varying backgrounds with special symmetries, by use of the Schwinger-Keldysh formalism. In a noninteracting theory the coherence information is shown to be encoded in new singular shells for the 2-point function. Imposing this phase space structure to the interacting theory leads to a a self-consistent equation of motion for a physcial density matrix, including coherence and a well defined collision integral. The method is applied \emph{e.g.~}to demonstrate how an initially coherent out-of-equlibrium state approaches equlibrium through decoherence and thermalization.
\end{abstract}


\end{frontmatter}

\section{Introduction} 
\label{intro}

Many problems in modern particle physics and cosmology require setting up transport equations for relativistic quantum systems in out-of-equilibrium conditions, including baryogenesis~\cite{BG,ClassForce,SemiClassSK}, heavy ion collisions~\cite{berges} and out-of equilibrium particle production~\cite{brandenberger}. It is straightforward to write a formal solution for the problem in the Shcwinger-Keldysh (SK) method, and detailed studies have been done on \emph{e.g.}  thermalization~\cite{generic_therm} of quantum systems. It is much more difficult to find a simple enough an approximation scheme that is usable in practical applications. Here we report a recent progress~\cite{HKR1,HKR2} towards a scheme, in which one can account for the nonlocal coherence effects in the presence of hard collisions. The key observation is that in cases with particular symmetries (the homogenous time dependent case and stationary, planar symmetric case) the free 2-point Wightmann functions have new, hitherto unnoticed singular shell solutions (at $k_0=0$ for the homogenous system and at $k_z=0$ in the static planar case), that carry the information about the nonlocal quantum coherence. When this shell structure is fed into the dynamical equations for the full interacting theory, they reduce to a closed set of quantum transport equations for the weight functions of the singular shells. These functions correspond to the on-shell particle numbers (mass shells) and to a measure of the coherence (the new $k_{0,z}=0$-shells). We have demonstrated the correctness of our interpretation of the coherence shells through application to the Klein problem~\cite{HKR1}. As an example of a case with collisions we have studied \emph{e.g.}~coherent production of unstable particles~\cite{HKR2}. Here we will show how an initially highly coherent, out-of-equilibrium quantum field configuration approaches equilibrium through decoherence and thermalization. We will mainly be concerned with relativistic fermions, but the method is applicable also to scalar fields and nonrelativistic problems.

\section{Free theory and the shell structure}
\label{free}
Let us consider the SK-equations for a collisionless fermion field, interacting with the background through a nontrivial mass term:
\beq
{\cal L} = i\bar \psi \deldag \psi
                    + \bar \psi_L m(t,\vec x) \psi_R
                    + \bar \psi_R m^*(t,\vec x) \psi_L \,. 
\label{freeLag1}
\eeq
In order to study the phase space structure of the correlator $iG^<(x,y) \equiv \langle \bar \psi (y)\psi (x) \rangle$, we write the SK-equations in the mixed representation. In the collisionless case, and to the lowest order in gradients the full SK-equation~\cite{SemiClassSK,HKR2} reduces to $(\kdag + \frac{i}{2} \deldag_x - m_R - i m_I \gamma^5) G^< = 0$. Consider first the homogenous case. Observing that helicity is a good quantum number, we can use the decomposition  $iG_h^<\gamma^0  \equiv g_h^< \otimes \frac12(1 + h \hat k\cdot \vec \sigma)$ in the chiral representation. Equation of motion for $g_h^<$ then breaks into hermitean and antihermitean parts:
\beq
{\rm (H)}:\quad 2 k_0 g_h^< = \{ H,g_h^< \}\,, \qquad
{\rm (AH)}:\quad i\partial_t g^<_h = [H, g^<_h] \,.
\label{Heq}
\eeq
\begin{figure}
\centering
\hskip -3truecm
\includegraphics[width=13cm]{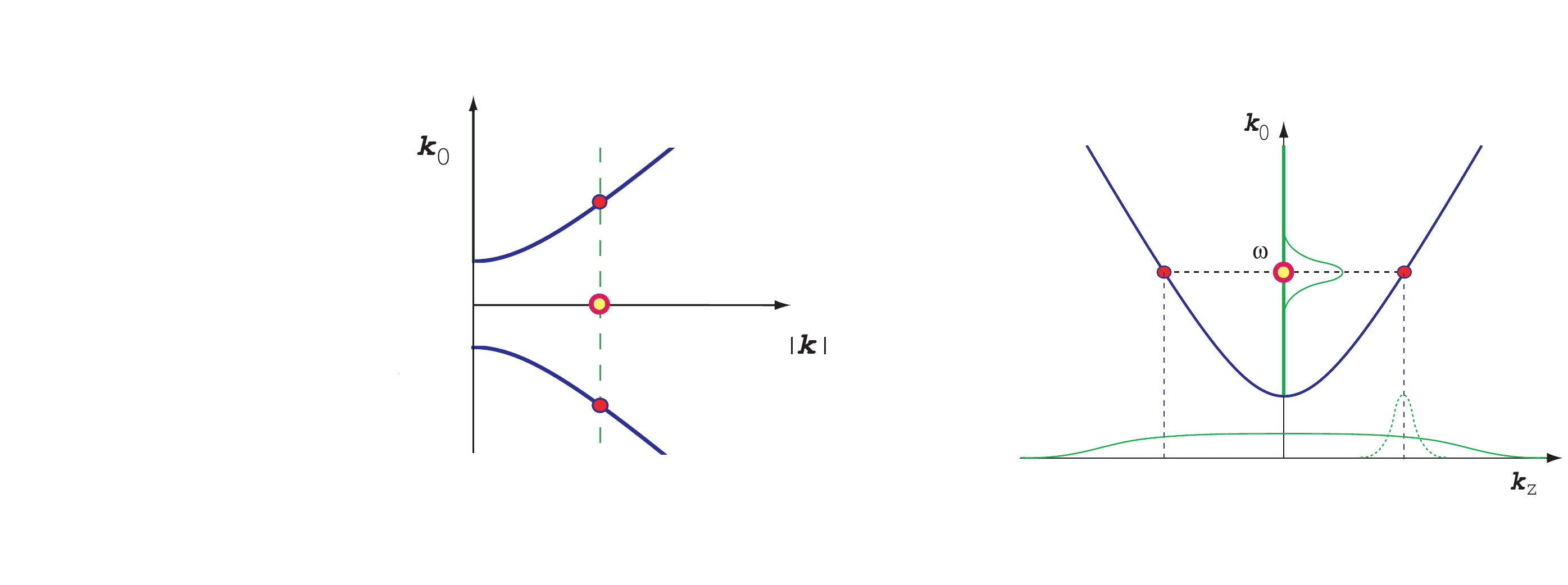}
\vskip -0.7cm 
    \caption{Dispersion relations in the case of a scalar mass function. 
    The dark filled (red) blobs show the mass-shell contributions and the light        (yellow) blobs show the corresponding coherence solution from the new $k_0=0$-shell in the homogenous case (left) and from the $k_z=0$-shell in the planar symmetric case (right).}
    \label{fig:DR-omega}
\end{figure}
where $H = - h |\vec{k}| \sigma_3 + m_R \sigma_1 - m_I \sigma_2$ is the local Hamiltonian. The H-equation is algebraic, and it constraints the form of the possible solutions. Using Bloch representation $2g_h^< = g^h_0 +  \vec g^h \cdot \vec \sigma$, it can be written as $B_{ab}g^h_b = 0$. This equation has nontrivial solutions if $
\det(B) = ( k_0^2 - \vec{k}^2 - |m|^2 )k_0^2 = 0$. Thus, somewhat surprisingly, in addition to the usual mass-shell solutions, there are new solutions at $k_0=0$ (see Fig.~(\ref{fig:DR-omega})). Explicit solutions on these shells are easily constructed:
\beqa
g^<_{h,{\rm m-s}}  &=& 
2 \pi {\rm s}_{k_0} \,f^h_{s_{k_0}} 
 \big(k_0 - h|{\bf k}|\sigma_3 + m_R \sigma_1 - m_I\sigma_2 \big) \delta (k^2-|m^2|)
\nonumber \\
g^<_{h,{\rm 0-s}} &=&  (\pi/|{\bf k}|) 
      \big(h\sigma_3 (m_R f^h_1 - m_I f^h_2) + |{\bf k}|(\sigma_1 f^h_1 + \sigma_2 f^h_2)\big) \; \delta(k_0) \,,
\eeqa
where $s_{k_0} \equiv {\rm sgn}(k_0)$ and the mass- and coherence shell factors $f^h_{s_{k_0}}$ and $f^h_{1,2}$ are some unknown functions of $t$ and ${\bf k}$. A similar construction can be done for the stationary, but planar symmetric case~\cite{HKR1}. In that case the spin $s$ orthogonal to the symmetry plane is conserved, and eventually, in a special frame, one finds equations very similar to Eqs.~(\ref{Heq}):
\beq
-2 s k_z g^<_s = P g_s^< + g_s^< P^\dagger \,, \qquad
is \partial_z g^<_s = P g_s^< - g_s^< P^\dagger \, ,
\label{Peq}
\eeq
where $P \equiv k_0 \sigma_3  + i(m_R \sigma_2 + m_I\sigma_1)$ is the local mean field momentum operator. The algebraic equation again gives rise to a singular shell structure; only this time the mass-shells are accompanied by a singular shell at $k_z=0$~\cite{HKR1}. The crucial issue leading to these new solutions is that \emph{we did not assume translational invariance in our treatment.} If one imposes translational invariance, the dynamical equations in (\ref{Heq}) and (\ref{Peq}) become algebraic as well and exclude the $k_{0,z}=0$-solutions.

The dynamical equations in (\ref{Heq}) and (\ref{Peq}) look compelling, but actually are meaningless because of the singular structure imposed by the algebraic constraints. Thus $g^<_{h,s}$'s are just phase space densities. Physical density matrices related to observable quantities can be introduced by convolving $g^<_{h,s}$'s with $k_0$- and/or $k_z$-dependent weight functions  that quantify the prior extrenous information on the system~\cite{HKR1}. Different types of such functions have been formally depicted in the right panel of Fig.~\ref{fig:DR-omega} by dashed lines. In a typical application, say, to the early universe there is no prior information at all, and physically meaningful equations are obtained by integrating the singular equations of motion with a flat weight. 

The scattering off a step potential (the Klein problem) provides a nice test bench for our formalism because there, in the collisionless limit, equations (\ref{Peq}) are {\em exact} equations for the problem. We applied eqs.~(\ref{Peq}) to the Klein problem in ref.~\cite{HKR1}, and our results demonstrated that the $k_z=0$-solutions are physical and describe the coherence: with them correct results for the transmission and reflection coefficients and tunneling factors follows, while neglecting them completely destroys the quantum nature of the problem.

\section{Interacting fields}
\label{intescalar}
\begin{figure}
\centering
\includegraphics[width=8cm]{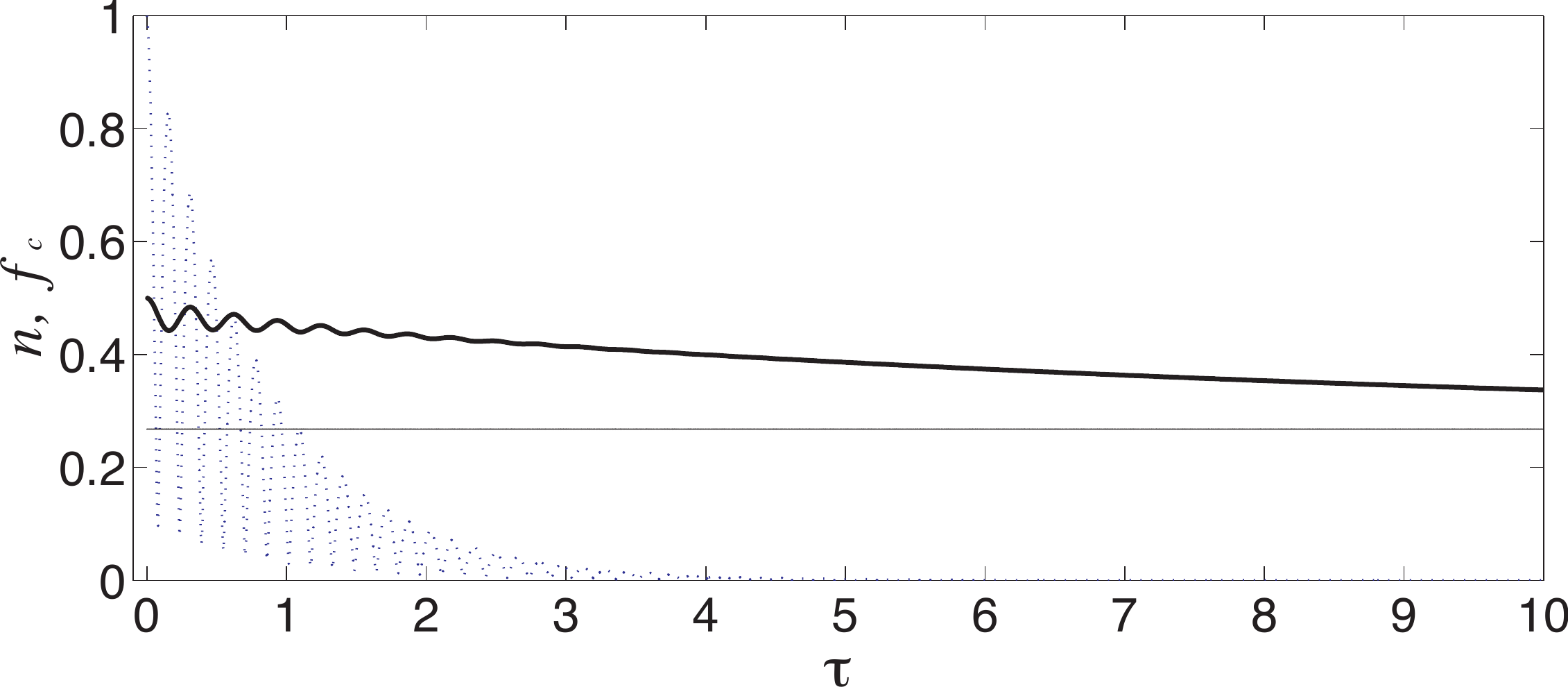}
\vskip -0cm 
    \caption{Evolution of the particle number $n_{\bf k} = f_+(|{\bf k}|)$    
    (solid line) 
    and the coherence $f_c \equiv (f_{1}^2 + f_{2}^2)^{1/2}$ (dashed line)
    for an initially highly coherent out-of-equilibrium state under decohering   
    interactions~\cite{HKR2}.}
    \label{fig:decoherence}	
\end{figure}

With interactions included the full Schwinger-Keldysh equations become much more complicated of course. A tractable approximation scheme can still be obtained by combining the familiar quasiparticle approximation with our mean field limit treatment of the constraint equations. Eventually the approximation is that the constraint equations are corrected by inclusion of the real parts of the self energies, while the collision terms affect only the evolution equations. Note that these are just the standard assumptions that one makes when deriving the usual Boltzmann equation from the SK-equations~\cite{Henning}. Here, for simplicity, we neglect also the real part of the self energy. Furthermore, if we compute collision terms using thermal distribution functions, the generalization of eqs.~(\ref{Heq}) to the interacting case can be written in the following form:
\beqa
{\rm (H):} && \quad 2k_0 g^<_h = \{H, g^<_h\} 
\nonumber \\
{\rm (AH):} &&\quad \partial_t g^<_h = -i[H, g^<_h] - \left\{D , \, g^<_h - (g^<_h)_{\rm
    eq} \right\} \,.
\label{rho_collHOMOG2}
\eeqa
where the precise form of the collision term $D$ depends on the interactions. When the evolution equation is integrated over the $k_0$-variable with a flat weight, contributions from each of the shells depicted in the left panel of the Fig.~2 contribute to the physical density matrix $\rho$. The evolution equation can be written in the form
\beq
\partial_t \rho_h = -i[H, \rho_h] - I_g(\rho_h) \,,
\label{rho_coll_intHOMOG}
\eeq
where the commutator term describes the usual quantum mechanical evolution of the density matrix, and the integral $I_g$ describes the effect of collisions. (For an explicit form of $I_g$ for a particular example see ref.~\cite{HKR2}.) The four independent components of $\rho_h$ are in one-to-one correspondence with the on shell functions $f^h_a$, $a = \pm, 1,2$. This allows for the collision integral to be evaluated in terms of $\rho_h$, leading to closure as is formally indicated in eq.~(\ref{rho_coll_intHOMOG}). It is important to note that the collision integral $I_g$ is nontrivial; it is computed at off shell momentum for the coherence shell. As a sample application of our method, we show in fig.~\ref{fig:decoherence} how an initially highly coherent out-of-equilibrium quantum state relaxes to equilibrium as a result of collisions~\cite{HKR2}.

Our method can also be applied to the bosonic fields with similar results. The coherence again resides at $k_{0,z}=0$-shells and a closed set of moment equations that include coherence and collisions can be found~\cite{HKR3}. It is also easy to write down similar quantum transport equations in the case of nonrelativistic fields~\cite{HKR3}.

\end{document}